\documentclass[usenatbib]{mn2e}
\usepackage[dvips]{graphicx}
\def\CN2{\mbox{$C_N^2 \ $}}

\title[]{Mesoscale optical turbulence simulations at Dome C: refinements}
\author[F. Lascaux et al.]{F. Lascaux,$^{1}$\thanks{E-mail:
     lascaux@arcetri.astro.it; masciadri@arcetri.astro.it} E. Masciadri$^1$\footnotemark[1],
and S. Hagelin$^{1, 2}$ \\ $^1$INAF Osservatorio Astrofisico
di Arcetri, Largo Enrico Fermi 5, I-501 25 Florence, Italy\\
$^2$Uppsala Universitet, Department of Earth Sciences, Villav\"agen 16,
S-752 36 Uppsala, Sweden}

\begin{document}

\newcommand{\cn}{$C_N^2$}

\label{firstpage}
\date{Accepted 2009 ??? ??, Received 2009 ??? ??; in original form
2009 ??? ??}  
\pagerange{\pageref{firstpage}--\pageref{lastpage}}
\pubyear{2008}

\maketitle

\begin{abstract}
In a recent paper the authors presented an extended study aiming at simulating the classical meteorological parameters 
and the optical turbulence at Dome C during the winter with the atmospherical mesoscale model Meso-NH. 
The goal of that paper was to validate the model above Dome C with the support of measurements and to use it afterwards above the Internal Antarctic Plateau to discriminate between the qualities of different potential astronomical sites on the plateau. A statistical analysis has been presented and the conclusions of that paper have been very promising. Wind speed and temperature fields (important for the computations of the optical turbulence parameters) revealed to be very well reconstructed by the Meso-NH model with better performances than what has been achieved 
with the European Centre for Medium-Range Weather Forecast (ECMWF) global model, especially near the surface. All results revealed to be resolution-dependent and it has been proved that a grid-nesting configuration (3 domains) with a high horizontal resolution ($\Delta$X = 1km) for the 
innermost domain is necessary to reconstruct all the optical turbulence features with a good correlation to measurements. 
High resolution simulations provided an averaged surface layer thickness just $\sim$14 m higher than what is estimated by measurements, the 
seeing in the free atmosphere showed a dispersion from the observed one of just a few hundredths of an arcsecond ($\Delta$$\varepsilon$ $\sim$ 0.05").
The unique limitation of the previous study was that the optical turbulence in the surface layer appeared overestimated by the model in both low and high resolution modes.
In this study we present the results obtained with an improved numerical configuration.
The same 15 nights have been simulated, and we show that the model results now match almost perfectly 
the observations in all their features: the surface thickness, the seeing in the free atmosphere as well as in the surface layer. 
This result permits us to investigate now other antarctic sites using a robust numerical model well adapted to the extreme polar conditions (Meso-NH).
\end{abstract}

\section{Introduction}

The ability of the model to forecast the evolution of the atmosphere and the optical turbulence
above the Antarctic Plateau has been extensively discussed in a previous paper by our team 
\citep{lf09}. In that paper all the 15 winter nights, for which measurements of the $\CN2$ are available (Trinquet et al. 2008) above Dome C, have been simulated with Meso-NH. 
The main conclusion of \cite{lf09} was that Meso-NH was able to reconstruct the optical turbulence in a region 
with extreme meteorological conditions such as Dome C. However a mono-domain configuration with low resolution (100 km) is not suitable to 
provide optical turbulence features well correlated to measurements while a grid-nesting configuration done with three domains 
(horizontal resolutions of 25 km, 10 km and 1 km) does it. More precisely, the mean simulated surface layer thickness $h_{sl,mnh-high}$=48.9 m $\pm$ 7.6 m was only 14 m higher than the observed one ($h_{sl,obs}$= 35.3 m $\pm$ 5.1 m). 
The median simulated free atmosphere seeing ($\varepsilon_{mnh,FA}$=0.35 $\pm$ 0.24 arcsec) was  
very well  correlated to the observed one ($\varepsilon_{obs,FA}$= 0.3 $\pm$ 0.2 arcsec). 
However the model tended to overestimate the intensity of the optical turbulence in the 
surface layer, thus generating a too strong median total seeing ($\varepsilon_{mnh,TOT}$= 2.29 $\pm$ 0.38 arcsec) if compared 
to the observed one ($\varepsilon_{obs,TOT}$=1.6 $\pm$ 0.2 arcsec).
This paper is intended to present the updated results for the same set of winter nights with an improved numerical configuration for 
the Meso-NH model. 
\begin{figure*}
\begin{center}
\includegraphics[width=0.8\textwidth]{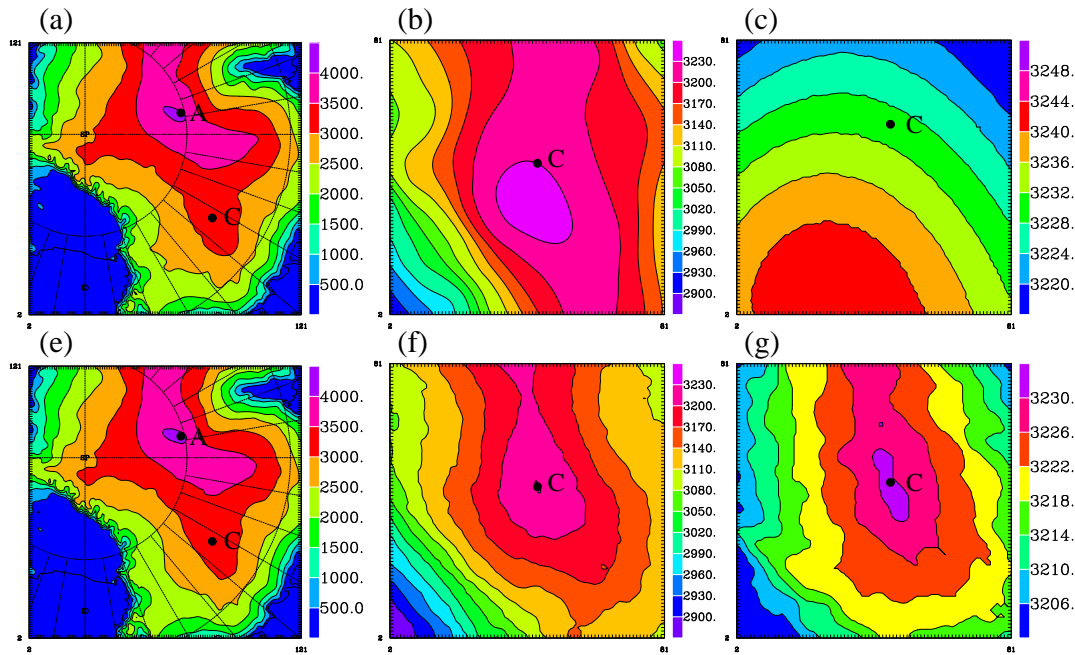}
\end{center}
\caption{Orography of Antarctica as seen by the Meson-Nh model (polar
 stereographic projection, grid-nesting configuration).
(a), (b) and (c) show the original orography (GTOPO30) with horizontal resolution of 25 km, 5 km and 1 km, respectively.
(d), (e) and (f) show the orography obtained with the new digital elevation model (RAMPDEMv2) with horizontal resolution of 25 km, 5 km and 1 km, respectively.
The dot labeled 'C' indicates the Concordia Station. The dot labeled 'A' indicates Dome A. The altitude is expressed in meter (m).}
{\label{fig_oro}}
\end{figure*}

\section{Improvements in the numerical configuration}
We refer the readers to \cite{lf09} for a complete overview of the numerical configurations of the mesoscale simulations. 
Here we briefly describe the differences in the numerical set-up with respect to the \cite{lf09} paper.\newline
{\bf (1)} A different digital elevation model (DEM), the so-called RAMPDEMv2 
(Radarsat Antarctic Mapping Project Digital Elevation Model, version 2 \citep{liu}) has been used instead of the GTOPO30 
DEM from the U.S. Geological Survey used in \cite{lf09}.
The improvement on the description of the orography is shown on Fig. \ref{fig_oro}.
The best description of the Dome C area is obtained with RAMPDEMv2, and is especially visible at horizontal resolutions below 5 km.
At $\Delta$X=5 km (Fig. \ref{fig_oro}b,f, with GTOPO30 and RAMPDEMv2 orographic models, respectively), one can see that the Concordia Station 
(labelled C on the figure) is located on the local summit with RAMPDEMv2, and is not with GTOPO30. 
This is even more evident inside the innermost domain ($\Delta$X=1 km, Fig. \ref{fig_oro}c,g).\newline  
{\bf (2)} The surface scheme of Meso-NH, the so called ISBA (Interaction Soil Biosphere Atmosphere) scheme, has been optimized for 
antarctic conditions: the thermal coefficient of the soil has been optimized for polar conditions
 and a climatological underground temperature T$_{c}$ has been introduced \citep{lem}. We briefly summarize the main concepts of that study. 
Our ability in well reconstructing the surface temperature T$_{s}$ is related to the ability in well reconstructing the sensible heat flux H 
that is responsible for the buoyancy-driven turbulence in the surface layer. 
The Meso-NH surface scheme is based on the force-restore method which consists in two equations that control the temporal evolution of the surface temperature T$_{s}$ and the deep temperature T$_{2}$ at a few tens of centimeters underground. The equation of the temporal evolution of the deep temperature has been modified adding a term depending on two free parameters: a climatological temperature T$_{c}$ and a relaxation term called $\gamma$. 
The two equations previously mentioned were forced by a one-year set of measurements of solar direct and long-wavelength radiations, the temperature T, wind speed V, pressure p and specific humidity q of the air above the ground.
The simulated T$_{s}$ and T$_{2}$ were compared to the observed one (i.e. the temperature measured at -5 cm and -30 cm from the ground) up 
to minimize the dispersion. 
This permitted to determine the two free parameters T$_{c}$ and $\gamma$.
Such a study permitted us to optimize the surface scheme ISBA for applications of the Meso-NH model to polar conditions.
\newline\newline
Coming back to our study, both mono-domain (low resolution) and grid-nesting (high resolution) configurations have been employed 
in \cite{lf09}. 
In this paper we could have restricted the study to the grid-nesting configuration (with high horizontal resolution) 
as it was the one giving the best results. 
However we decided to present also the results of the low horizontal resolution to confirm the previous conclusions 
of \cite{lf09} regarding the horizontal resolution necessary to be used for obtaining reliable forecasts.

\section{Optical Turbulence forecast}

We focus our attention on the prediction of the 3D maps of \cn~obtained with the Astro-Meso-NH package implemented by our team in 
the Meso-NH mesoscale model (Masciadri et al. (1999a), (1999b)) and validated in successive phases above different astronomical sites (among 
the most important Masciadri \& Jabouille (2001), Masciadri et al. (2004), Masciadri \& Egner (2006)). The model is employed here with the 
new configuration mentioned in the previous section.
We especially look at three different features that mainly characterize the optical turbulence, already discussed in \cite{lf09}:
\begin{itemize}
\item the surface layer thickness;
\item the median free atmosphere seeing;
\item the median total seeing (from the ground up to 10 km above ground level).
\end{itemize}
The study is performed for the same set of 15 winter nights used in \cite{lf09} and employing the same analysis.
We first present the results of prediction of the surface layer thickness and then we present the forecasting of the free 
atmosphere seeing ${\varepsilon}_{FA}$ and the total seeing ${\varepsilon}_{TOT}$.

\subsection{Surface layer thickness ($h_{sl}$)}
The same method used in \cite{lf09} is employed to determine the surface layer thickness of each nights. We highlight that 
this is not the only way to define the surface layer but we are forced to use this method employed by \cite{tr} 
to be able to compare simulations with measurements. The thickness $h_{sl}$ is defined as the vertical slab containing
90\% of the optical turbulence developed in the first kilometer:
\begin{equation}
 \label{eq:bl1}
 \frac{ \int_{8m}^{h_{sl}} C_N^2(h)dh }{ \int_{8m}^{1km} C_N^2(h)dh } < 0.90
\end{equation}
where $C_N^2$ is the refractive index structure parameter.
\begin{table}
 \centering
  \caption{Surface layer thickness $h_{sl}$ for 15 winter nights (Trinquet et al. 2008). 
           The mean value is reported with the associated standard deviation ($\sigma$) and statistical error ($\sigma$/$\sqrt{N}$) where N is the number of independent estimates that is independent nights. Units in meter (m).}
  \begin{tabular}{|c|c|c|c|}
   \hline
   Date &  Observed surface  & Date & Observed surface    \\
        & layer thickness &      & layer thickness \\
   \hline
   04/07/05 & 30 & 12/08/05 & 22 \\
   07/07/05 & 21 & 29/08/05 & 47 \\
   11/07/05 & 98 & 02/09/05 & 41 \\
   18/07/05 & 26 & 05/09/05 & 20 \\
   21/07/05 & 47 & 07/09/05 & 39 \\
   25/07/05 & 22 & 16/09/05 & 24 \\
   01/08/05 & 40 & 21/09/05 & 22 \\
   08/08/05 & 30 & \\
   \hline
            & Mean  & 35.3 & \\
   \hline
             & $\sigma$  & 19.9 & \\
   \hline
          & $\sigma$/$\sqrt{N}$  & 5.1 & \\
  \hline
  \end{tabular}
 \label{tab:hl0}
\end{table}

\begin{figure}
\begin{center}
\includegraphics[width=4.5 cm]{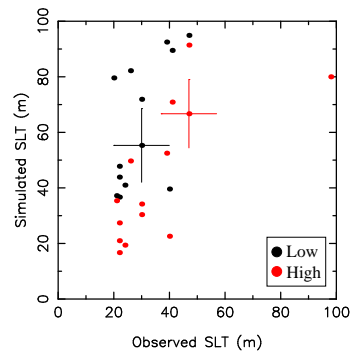}
\end{center}
\caption{Correlation plot between measured and simulated surface layer thicknesses (black: mono-domain
configuration; red: grid-nested configuration). For the simulated values only the mean values between 12 UTC
and 16 UTC are considered.
For each configuration of the simulation (high and low horizontal resolution) the error bars are
reported for one point only (and are equal to $\sigma$). Units are in meter (m).}
{\label{fig:corr_slt}}
\end{figure}

We report in Table \ref{tab:hl0} the values of the observed surface layer thicknesses and the corresponding mean and standard deviation ($\sigma$) and statistical error ($\sigma$/$\sqrt{N}$) for the 15 nights.
The simulated $h_{sl}$ for both mono-domain configuration (i.e. low horizontal resolution) and grid-nesting configuration 
(i.e. high horizontal resolution) are summarized on Table \ref{tab:hl1}.
The values related to each night as well as the mean $h_{sl,mnh-low}$ and $h_{sl,mnh-high}$ are computed.
A careful examination of the night of September 5, 2005 (for which the computed $h_{sl,mnh-high}$ was 338.4 m) 
highlighted the fact that the value of the surface layer retrieved with the criterion reported in Eq.\ref{eq:bl1} was unexploitable for this night. 
Indeed a look at the temporal evolution of the \cn~profile during that night (not shown here)  
clearly indicates a very weak turbulent night, with maxima of \cn~concentrated well below 30 m. Using the criterion expressed by Eq.\ref{eq:bl1} it follows that $h_{sl}$ has to be necessarily large to achieve the 90$\%$ of the turbulence developed in the first kilometer in this night.  
That's why we decided to remove this night from the mean computation in the high resolution case in order to have a more representative 
$h_{sl,mnh-high}$. Anyway, this removal does not affect the qualitative conclusions inferred from the comparison between the observed 
and the computed mean values.
\par
The high resolution configuration gives a mean surface layer thickness $h_{sl,mnh-high}$ = 44.2 $\pm$ 6.6 m (with $\sigma$ = 24.6 m)
and the low resolution configuration a $h_{sl,mnh-low}$ = 68.4 $\pm$ 6.8 m (with $\sigma$ = 26.5 m).
We confirm, therefore, the same conclusion obtained in \cite{lf09}: the surface layer thickness computed in grid-nested mode is closer to the observed one ($h_{sl,obs}$ = 35.3 $\pm$ 5.1m). 
Figure \ref{fig:corr_slt} shows the correlation between observed and simulated surface layer thicknesses with the associated intrinsic dispersion (error bar) equal to the $\sigma$ values. 
This figure confirms the tendency of the high resolution simulations to give $h_{sl}$ values closer to the observed ones from a statistical point of view.
One can see that the dispersion of the simulated values (${\sigma}_{mnh-high}$ = 24.6 m and ${\sigma}_{mnh-low}$ = 26.5 m) are just slitghly 
higher than the observed ${\sigma}_{obs}$ = 19.9 m.  Also the dispersion of the surface layer thicknesses is, therefore, reasonably well reconstructed by the model. If we take into account the statistical error in the high resolution case ($\sigma$/$\sqrt{N}$), we conclude that the mesoscale model provides, for this statistical sample, a typical surface-layer thickness just $\sim$ 3 m higher than the observed one. We have therefore an excellent estimated mean, almost within the statistical uncertainty.

\subsection{Optical turbulence vertical distribution: seeing in the free atmosphere and in the whole atmosphere}
\begin{table}
 \caption{Mean surface layer thickness $h_{sl}$ for the same 15 winter nights that are reported in Table
          \ref{tab:hl0}, deduced from Meso-NH computations using the criterion in
          Eq. \ref{eq:bl1} in the temporal range 12:00-16:00 UTC. The mean value is reported  with the associated statistical error $\sigma$/$\sqrt{N}$. Units in meter (m). }
          \begin{tabular}{|c|c|c|}
 \hline
 \multicolumn{1}{|c|}{Date} &
 \multicolumn{1}{c|}{Surface layer thickness} &
 \multicolumn{1}{c|}{Surface layer thickness} \\
 \multicolumn{1}{|c|}{} &
 \multicolumn{1}{c|}{Meso-NH grid-nesting} &
 \multicolumn{1}{c|}{Meso-NH mono-domain} \\
 \hline
 04/07/05 & 30.4 & 55.3 \\
 07/07/05 & 35.4 & 37.2 \\
 11/07/05 & 80.0 &108.3 \\
 18/07/05 & 49.7 & 82.2 \\
 21/07/05 & 66.7 & 94.9 \\
 25/07/05 & 27.4 & 36.7 \\
 01/08/05 & 22.6 & 39.6 \\
 08/08/05 & 34.2 & 71.9 \\
 12/08/05 & 16.7 & 43.9 \\
 29/08/05 & 91.4 &106.1 \\
 02/09/05 & 70.9 & 89.5 \\
 05/09/05 &338.4 & 79.6 \\
 07/09/05 & 52.5 & 92.5 \\
 16/09/05 & 19.4 & 41.0 \\
 21/09/05 & 21.0 & 47.8 \\
 \hline
 Mean     & 44.2* & 68.4\\
 \hline
 $\sigma$    & 24.6* & 26.5 \\
 \hline
 $\sigma$/$\sqrt{N}$   &  6.6* &  6.8  \\
 \hline
\multicolumn{3}{|c|}{*These values are computed without taking into account the} \\
\multicolumn{3}{|c|}{night of the 05/09/05 (see text for further explanations).}
\end{tabular}
 \label{tab:hl1}
\end{table}

\begin{table}
\caption{Total seeing $\varepsilon_{TOP}$$=$$\varepsilon_{[8m,h_{top}]}$ and seeing in the free atmosphere 
$\varepsilon_{FA}$$=$$\varepsilon_{[h_{sl},h_{top}]}$ calculated for the 15 nights and averaged in the temporal range 
12:00-16:00 UTC. 
See the text for the definition of h$_{sl}$ and h$_{top}$. 
In the second column are reported the observed values, in the third and fourth columns the simulated values 
obtained with high and low horizontal resolution respectively. Units in arcsec.}
\begin{tabular}{cccc}
\hline
& Obs.        & MESO-NH              & MESO-NH            \\
&             & HIGH                 & LOW                \\
\hline                                       
  Date     & $\varepsilon_{FA}$/$\varepsilon_{TOT}$      & $\varepsilon_{FA}$/$\varepsilon_{TOT}$      & $\varepsilon_{FA}$/
$\varepsilon_{TOT}$        \\
           & {\tiny($h_{sl}$=33m)} & {\tiny($h_{sl}$=44.2m)} & {\tiny($h_{sl}$=62.4m)} \\
 \hline
  04/07/05 &      0.3 / 1.6       &     0.22 / 2.28      &    0.35 / 2.13         \\
  07/07/05 &      0.2 / 1.5       &     0.28 / 1.91      &    0.24 / 1.49         \\
  11/07/05 &      1.4 / 1.7       &     1.61 / 1.81      &    1.57 / 2.08         \\
  18/07/05 &      0.3 / 2.0       &     0.80 / 1.94      &    1.98 / 2.20         \\
  21/07/05 &      0.7 / 1.1       &     0.86 / 1.27      &    0.41 / 1.02         \\
  25/07/05 &      0.3 / 1.0       &     0.25 / 0.85      &    0.23 / 2.03         \\
  01/08/05 &      0.8 / 1.6       &     0.22 / 2.27      &    0.22 / 2.05         \\
  08/08/05 &      0.5 / 2.3       &     0.35 / 1.70      &    0.31 / 1.00         \\
  12/08/05 &      0.2 / 1.5       &     0.23 / 0.99      &    0.28 / 2.49         \\
  29/08/05 &      2.5 / 3.6       &     2.29 / 2.47      &    2.12 / 2.78         \\
  02/09/05 &      0.9 / 1.9       &     1.16 / 1.54      &    0.92 / 1.77         \\
  05/09/05 &      0.3 / 1.0       &     0.30 / 0.52      &    0.27 / 0.73         \\
  07/09/05 &      1.4 / 2.8       &     1.69 / 3.73      &    1.88 / 3.69         \\
  16/09/05 &      0.2 / 1.5       &     0.21 / 1.57      &    0.20 / 2.25         \\
  21/09/05 &      0.2 / 1.7       &     0.26 / 1.63      &    0.22 / 0.75         \\
 \hline
  Median   &      0.3 / 1.6       &     0.30 / 1.70      &    0.31 / 2.05         \\
 \hline
  $\sigma$ &      0.7 / 0.7       &     0.67 / 0.77      &    0.66 / 0.81         \\
 \hline
  $\sigma$/$\sqrt{N}$ & 0.2 / 0.2 &     0.17 / 0.21      &    0.17 / 0.21         \\
 \hline
\end{tabular}
\label{see}
\end{table}

\begin{figure}
\begin{center}
\includegraphics[width=4.5 cm ]{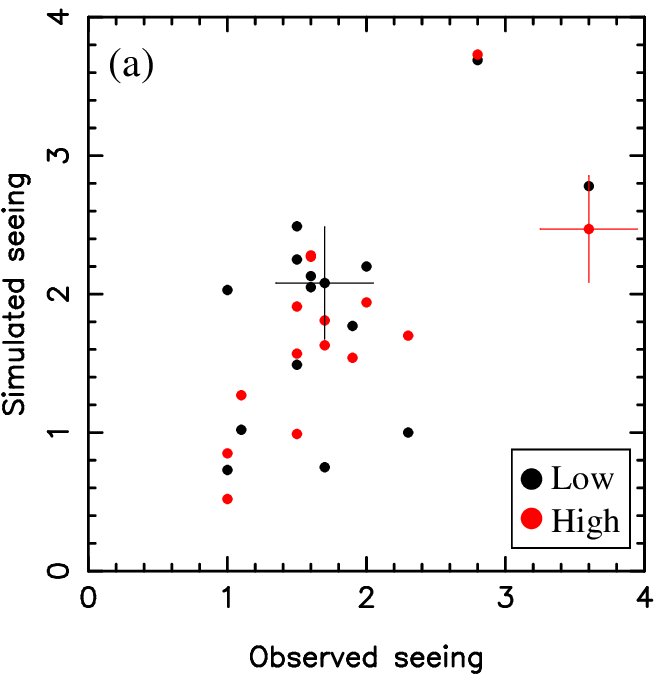}
\includegraphics[width=4.5 cm]{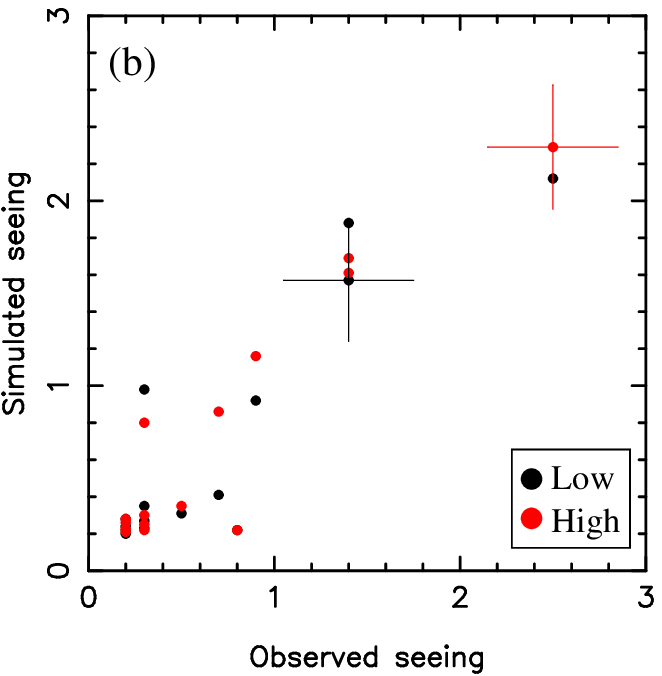}
\end{center}
\caption{Correlation plot between measured and simulated (a) total seeing  and (b) free atmosphere seeing.
Black dots: low resolution - Red dots: high resolution.
The error bars for simulations and observations are reported for one point only (and are equal to $\sigma$). Units are in arcsec.}
{\label{fig:corr_see}}
\end{figure}

\begin{figure}
\begin{center}
\includegraphics[width=7.0 cm]{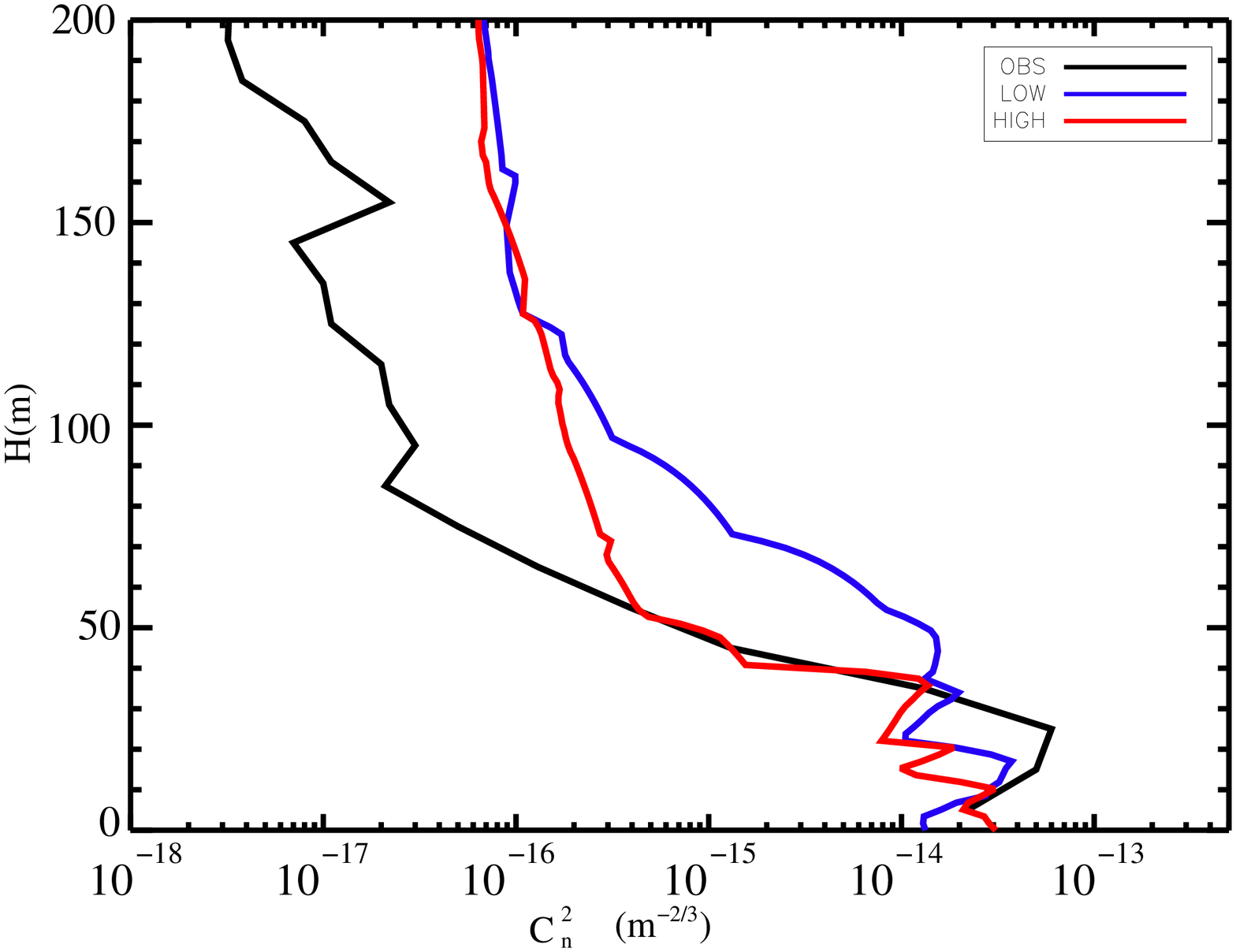}
\includegraphics[width=7.0 cm]{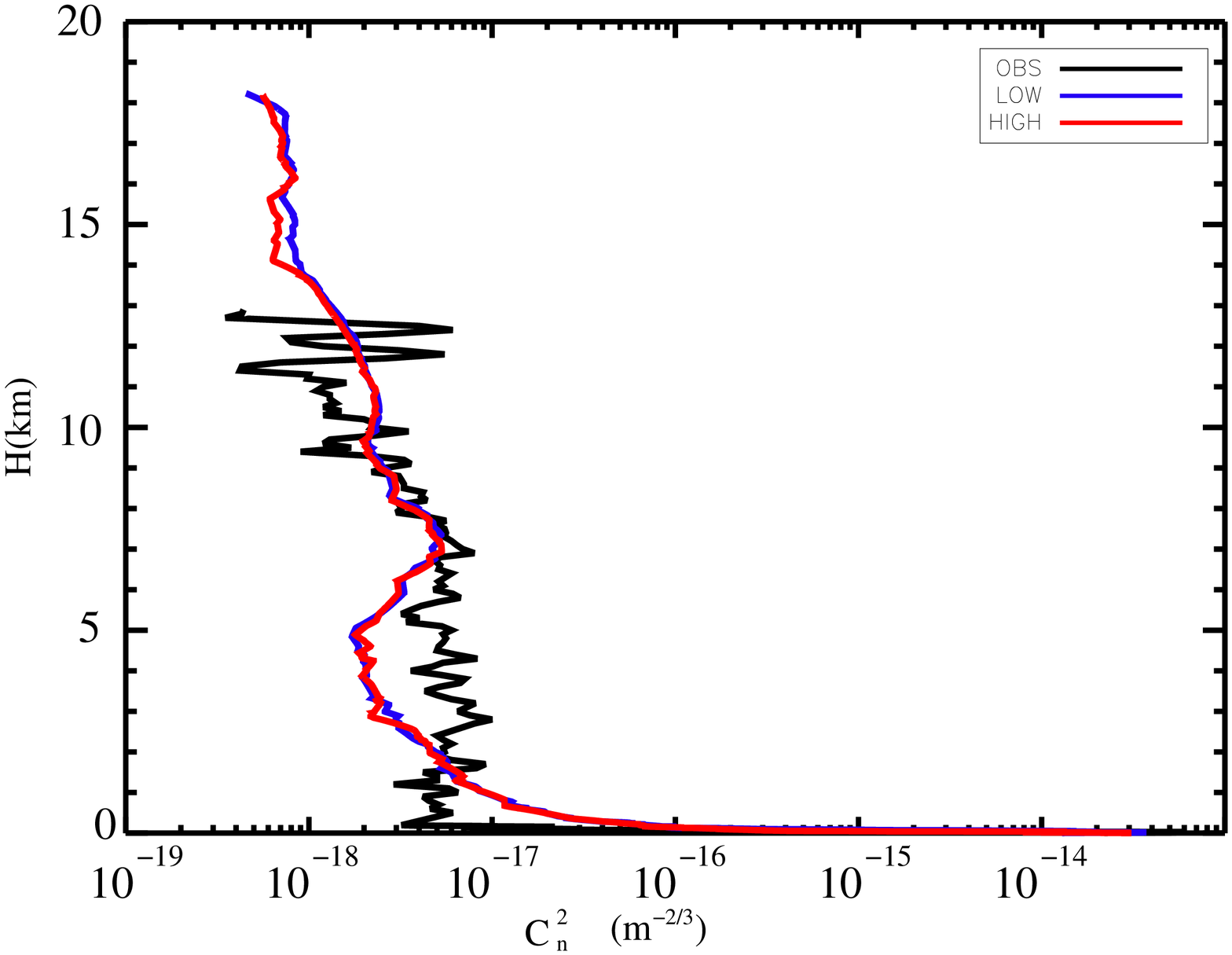}
\end{center}
\caption{Median \cn~profile measured (black line) with microthermal sensors mounted 
on balloons (from Trinquet et al. 2008) and simulated with the Meso-NH mesoscale model with the 
low-horizontal resolution (blue line) and the high-horizontal resolution (red line), computed over the 15 nights. 
Up: from the ground up to 200 m. Down: from the 
ground up to 20 km. Simulations are considered in the temporal range 12:00-16:00 UTC. Units in m$^{-2/3}$.}
{\label{pv_cn2}}
\end{figure}

Table \ref{see} reproduces the observed and simulated total seeing ($\varepsilon_{TOT}$) and free atmosphere seeing ($\varepsilon_{FA}$) for each night. The median values with the respective standard deviation ($\sigma$) and statistical error ($\sigma$/$\sqrt{N}$) are reported for the low and high resolution cases. One can see that the median total seeing calculated with the low resolution mode ($\varepsilon_{TOT,mnh-low}$ = 2.05 
$\pm$ 0.21 arcsec) still remains larger than the observed one ($\epsilon_{TOT,obs}$ = 1.6 $\pm$ 0.2 arcsec) as it has been observed in \cite{lf09}.
However the median forecasted seeing in the high resolution mode matches almost perfectly the observations ($\varepsilon_{TOT,mnh-high}$ = 1.7 
$\pm$ 0.21 arcsec). The same can be stated for the median free atmosphere seeing reconstructed by the model ($\varepsilon_{FA,mnh-high}$ = 0.30 $\pm$ 0.17 arsec) that is very well correlated to the observed one ($\epsilon_{FA,obs}$ = 0.30 $\pm$ 0.20 arcsec).
Figure \ref{fig:corr_see} shows the correlation plots for the total and free atmosphere seeing, for each of the 15 investigated winter nights. 
This figure confirms the good agreement between model and observation from a statistical point of view. 
How do the modifications implemented in the Meso-NH model change the the vertical distribution of the optical turbulence (\cn~profiles)?
Figure \ref{pv_cn2} shows the vertical profiles of the observed and simulated \cn~profiles obtained with the low and high resolution modes. 
We can observe that, similarly to what has been found in \cite{lf09}, near the ground the high resolution model provides a sharper decrease 
in the optical turbulence than the low resolution mode. 
However, with this new model configuration, the shape of the \cn~profile is better correlated to measurements in the first 60 m than with the 
previous configuration. 
We finally observe that, in both cases, there is
still space for future improvements in some vertical slabs.
We expressly avoid more complex calibration specific for a
particular site as that presented in \cite{m3},
because we intend to use the same model to discriminate
the quality of other sites above the internal plateau.

\section{Conclusions}
In this paper simulations of the \cn~profiles related to all the 15 nights for which measurements done at Dome C in winter are available, have been compared to measurements.
We conclude, that, in the present configuration, the Meso-NH model with the 'Astro-Meso-NH' package provides excellent estimates of the optical turbulence at Dome C, from a qualitative as well as quantitative point of view, if used in the high resolution mode.  All the conclusions achieved in \cite{lf09} remains unchanged. 
However we clearly put in evidence, in this paper, the positive impact of the new numerical configuration of the model
Meso-NH on the forecast of the optical turbulence, especially for the surface layer thickness and the total and 
free atmosphere seeing. In synthesis, the Meso-NH model provides:\newline
{\bf (1)}  a simulated mean surface layer thickness $h_{sl,mnh-high}$ = 44.2 $\pm$ 6.6 m versus the observed $h_{sl,obs}$ = 35.3 $\pm$ 5.1 m. \newline
{\bf (2)}  a simulated median total seeing $\varepsilon_{TOT,mnh-high}$ = 1.7 $\pm$ 0.21 arcsec versus the observed $\varepsilon_{TOT,obs}$ =  1.6 $\pm$ 0.2. \newline
{\bf (3)}  a simulated median free atmosphere seeing $\varepsilon_{FA,mnh-high}$ = 0.30 $\pm$ 0.17 arcsec versus the observed $\varepsilon_{FA,obs}$ =  0.30 $\pm$ 0.20. \newline
\par
We are now ready to apply this model to other antarctic regions and to explore the optical turbulence features on others potential interesting sites 
such as Dome A, South Pole, or even Ridge A, in order to perform sites inter-comparison and identify the best location for astronomical applications.

\section*{Acknowledgments}
This study has been funded by the Marie Curie Excellence Grant (FOROT) - MEXT-CT-2005-023878. The Meso-NH model is initialized with European Centre for Medium-Range Weather Forecasts (ECMWF) GRIB files. Access to ECMWF products is authorized by the Meteorologic Service of the Italian Air Force. We also thank the CNRM-LA Meso-NH user support team.


\label{lastpage}
\end{document}